# Field-Induced Up–Up–Down State and Frustrated Magnetism in a Non-Kramers Triangular Antiferromagnet

Zhaoyi Li[1,2,5], Qinchen Duan[3,4], Bo Wen[5], Ruidan Zhong[3,4], Shu Guo[1,2,*]

[1]Shenzhen Institute for Quantum Science and Engineering, Southern University of Science and Technology, Shenzhen 518055, China.

[2]International Quantum Academy, Shenzhen 518048, China.

[3]Tsung-Dao Lee Institute, Shanghai Jiao Tong University, Shanghai 201210, People's Republic of China.

[4]School of Physics and Astronomy, Shanghai Jiao Tong University, Shanghai 200240, People's Republic of China.

[5]School of Physics and Electronics, Henan University, Kaifeng 475004, China.

**Abstract**

A previously unreported triangular lattice (TL) antiferromagnet, $TmZnGaO_4$, was synthesized as single crystals, and its crystal structure, magnetic susceptibilities, and specific heat were reported. Its crystal structure is isomorphic to that of the transverse-field Ising antiferromagnet $TmMgGaO_4$, with $Tm^{3+}$ ions located in the TLs, separated by a nonmagnetic bilayer composed mainly of $Ga^{3+}$ and $Zn^{2+}$ ions. The magnetic susceptibilities indicate the dominating antiferromagnetic interactions. The magnetization curves (*M*–*H*) exhibit strong easy-*c*-axis anisotropy, with a clear one-third magnetic plateau emerging, consistent with a field-induced up–up–down (↑↑↓) spin configuration. Instead of forming a conventional long-range magnetic order, the system exhibits two broad anomalies at 0.11 K and 2.81 K in zero-field specific heat measurements, highlighting the persistence of strong spin fluctuations and the potential for exotic quantum spin states. The above results reveal its future interest in exploring exotic quantum spin states in $TmZnGaO_4$.

## 1. Introduction

In frustrated magnetic systems, particle interactions are constrained by geometric or competing exchange interactions, preventing them from attaining the classical minimum-energy configuration.[1,2] In triangular-lattice (TL) antiferromagnets, the nearest-neighbor (NN) isotropic Heisenberg model has been shown to form a stable non-collinear 120° magnetic structure.[3,4] However, theoretical studies have further shown that relying solely on NN exchange interactions is insufficient to describe the complex magnetic behavior in TL antiferromagnets. By introducing next-NN exchange interactions and magnetic anisotropy, competing magnetic interactions can be induced, leading to significant frustration, potentially suppressing long-range magnetic order (LRO), and promoting the emergence of unconventional quantum spin states. [5-7] In rare-earth (*RE*) TL magnets, the magnetic moments originate from the localized 4*f* electrons of $RE^{3+}$. The combined effect of strong spin-orbit coupling (SOC) and crystal electric field (CEF) results in the low-energy degrees of freedom of these systems usually exhibiting highly anisotropic effective spin, which typically deviates from the simple isotropic Heisenberg model.[6] The distinct 4*f* electron configurations give rise to a rich diversity of physical properties and highly tunable magnetic ground states. Especially in magnetic systems with small spin and reduced dimensionality, the cooperative effects of quantum fluctuations and geometrical frustration may lead to ground states beyond conventional magnetic orders, thereby providing abundant opportunities for experimental and theoretical exploration of novel quantum spin states, such as quantum spin liquids[8,9], magnetization plateaus[10,11], spin glass state[12], and multipolar excitation[13].

In *RE*-based frustrated magnets, the low-energy magnetic behavior is dominated by the lowest-lying states, which can be determined by the CEF, and distinct 4*f* electron configurations give rise to a rich diversity of physical properties and highly tunable magnetic ground states.[14,15] Most studies have focused on doublet systems that are protected by symmetry, where odd-electron Kramers ions depend on time-reversal symmetry (TRS)[16], while non-Kramers ions are mainly affected by local crystal symmetry.[10,17] Since non-Kramers ions are not protected by TRS, they are easily influenced by the local environment and are sensitive to perturbations that break local symmetry. As a unique class of low-dimensional frustrated magnets, non-Kramers $RE^{3+}$-based systems with a quasi-doublet ground state are expected to unveil exotic quantum frustration physics governed by local CEF symmetry rather than TRS.[18-20] For instance, in the Tm-based TL antiferromagnet $TmMgGaO_4$ (TMGO)[10,21-30], the $Tm^{3+}$ ($4f^{12}$) ion features strong SOC that combines its orbital ($L$ = 5) and spin ($S$ = 1) angular momenta into a total angular momentum of $J$ = 6. Under the splitting of the crystal field, the 13 degenerate states corresponding to it will undergo further energy level splitting. Studies have shown that the low-energy effective degrees of freedom in this ion system are a pair of quasi-degenerate singlets with extremely close energies. The two levels have a energy gap relative to the other crystal-field levels, forming a quasi-doublet ($S_{eff}$ = 1/2) that is well separated from the other levels.[23-25] At finite temperatures, it has been proposed that there may be a quasi-long-range Berezinskii-Kosterlitz-Thouless (BKT) phase with emergent U(1) symmetry.[25,26,31] Furthermore, the nuclear magnetic resonance (NMR) experiments have supported the existence of the above-mentioned BKT phase.[27] A BKT physics has also been proposed in the other $Tm^{3+}$-based TL antiferromagnet $TmZn_2GaO_5$.[32]

Here, $TmZnGaO_4$ (TZGO), an isostructural analog of TMGO, was synthesized via the high-temperature self-flux method. By replacing non-magnetic $Mg^{2+}$ with $Zn^{2+}$ while maintaining the integrity of the two-dimensional (2D) magnetic TLs, the interlayer coupling, lattice degrees of freedom, and crystal-field environment in the layered TL antiferromagnetic (AFM) system were regulated. Magnetic susceptibility measurements show an easy-*c*-axis magnetic anisotropy, and the negative Weiss temperature indicates that the system primarily exhibits AFM interactions. Under an applied magnetic field, the system exhibits a clear one-third magnetization plateau, indicating a field-induced up-up-down (UUD, ↑↑↓) spin configuration. Specific heat measurements above 50 mK reveal bulge anomalies at 0.11 K and 2.81 K in the non-Kramers *RE*-based TZGO. These results provide an experimental basis for exploring the microscopic mechanism of frustrated magnetism in this system.

## 2. Results and Discussion

### 2.1. Crystal structure

TZGO crystallizes in the trigonal crystal system with a space group *R*-3*m* (No. 166), with unit cell parameters $a$ = 3.43 Å and $c$ = 24.99 Å (**Figure 1a**). The magnetic $Tm^{3+}$ forms a TL within the *a*-*b* plane (**Figure 1b**), which is stacked in an ABCABC manner along the crystallographic *c*-axis (**Figure 1c)**, interconnected via double-layered triangular bipyramids, where $Ga^{3+}$ and $Zn^{2+}$ atoms statistically share the same crystallographic site (6*c*). This resulted in a larger interlayer spacing ($d_{\text{inter}}$ = 8.33 Å) than that of TL layers ($d_{\text{intra}}$ = 3.43 Å), effectively weakening the magnetic exchange interaction between neighboring layers and leading to a 2D magnetic framework. The inter-layer site-mixed Ga/Zn environments may split the Tm site into two positions displaced along the crystallographic *c*-axis, causing positional disorder of Tm within the $TmO_6$ octahedra (**Figure 1d**), a structural feature also observed in TMGO.[21] Millimeter-sized single crystals were obtained via the self-flux method (**Figure 1e**). Additionally, P-XRD was performed both on the (001) plane of the transparent millimeter-sized plate-like single crystal (**Figure 1f**) and on the powder obtained by grinding the single crystals (**Figure 1g**). The powder sample data were refined using GSAS-II[33], showing excellent agreement with the Rietveld refinement profile ($R_{\text{wp}}$ = 4.40% and $\chi^2$ = 3.30). The measurement results confirm the quality and purity of the crystals. The experimental diffraction data obtained from APEX4 match well with the theoretical values derived from SingleCrystal software[34], confirming the accuracy of the structure (**Figure 1h**).

### 2.2. Magnetism

Anisotropic magnetic susceptibility measurements were performed on the crystal with the magnetic field applied parallel to ($B_{\parallel}$) and perpendicular to ($B_{\perp}$) the TL layer. Magnetic susceptibilities were measured from 1.8 to 100 K under a magnetic field of 0.2 T along both $B_{\parallel}$ and $B_{\perp}$ directions (**Figure 2a**), and no LRO has been observed. The magnetic susceptibility in different orientations exhibits significant anisotropy, stemming from the highly anisotropic ground-state doublet induced by the strong CEF effect and the 2D crystal structure. For magnetic fields oriented along $B_{\parallel}$, the magnetic susceptibility deviates significantly from the standard Curie-Weiss (C-W) behavior (**Figure S1**). In contrast, for magnetic fields oriented along $B_{\perp}$, the magnetic susceptibility obeys C-W behavior. Such behavior is a typical feature of *RE*-based systems with strong crystal-field anisotropy.[10,20,21,35]

In the temperature range of 10–100 K for $B_{\perp}$, the $\chi^{-1}$ were fitted using the C-W law $\chi - \chi_0 = \frac{C}{T-\Theta}$ (where $\Theta$ is the C-W temperature and $C$ is the Curie constant, $\chi_0$ is the temperature-independent susceptibility). The fitting yielded $\chi_0$ = -0.0052 emu·mol$^{-1}$·Oe$^{-1}$, $\Theta$ = -18.65 K and $C$ = 14.71 emu·K·mol$^{-1}$·Oe$^{-1}$. The negative Weiss temperatures indicate predominant AFM interactions in TZGO. The effective magnetic moment for $Tm^{3+}$ ions was calculated using the equation $\mu_{\mathrm{eff}} = \sqrt{8C}\mu_{\mathrm{B}}$ ($\mu_{\mathrm{B}}$ is the Bohr magneton) to be $\mu_{\mathrm{eff}\perp}$ = 10.85 $\mu_{\mathrm{B}}$. The field dependence of the magnetization for $B_{\perp}$ and $B_{\parallel}$ measured at 2 K (**Figure 2b**) also exhibits a strong anisotropy with $M_{(\max,\perp)}$ ten times larger than that of $M_{(\max,\parallel)}$. At the same time, when the magnetic field is perpendicular to the *a*-*b* plane, a magnetic polarization intensity of approximately 6.50 $\mu_{\mathrm{B}}$ is observed, approaching the saturation value. Based on the saturation magnetic moment and assuming an effective pseudospin-1/2 ground state, the out-of-plane *g* factor $g_{\perp}$ is evaluated by the equation $M_{\mathrm{sat}} = g_{\perp}S_{\mathrm{eff}}\mu_{\mathrm{B}}$. From this relation, the value of $g_{\perp}/2$ is determined to be 6.50, which subsequently yields an effective $g_{\perp}$ of 13.00. This remarkably large value is highly comparable to that observed in TMGO.[10]

From magnetic susceptibility measurements with a $^3$He option down to 0.4 K (**Figure 3a**), a broad hump was observed without signs of clear LRO. The field dependence of magnetization from 0 to 7 T was measured at several temperatures below 2 K (**Figure 3b**). As the temperature decreased, a plateau anomaly emerged around 1.65 T. Differential analysis of the magnetization at 0.4 K revealed a minimum at 1.65 T, corresponding to a magnetic plateau (**Figure 3c**). This magnetic plateau accounts for approximately one-third of the saturated magnetic moments, suggesting a field-induced

UUD spin configuration.[10] The field dependence of magnetization was also measured for $B_{\parallel}$ directions below 2 K (**Figure 3d**), in which no anomaly was observed below 7 T.

**2.3. Specific heat**

To further investigate the magnetic ground states of TZGO, specific heat measurements are performed on single crystals in both zero field and various fields with $B_{\perp}$. The variation in specific heat ($C_p$) with temperature under different magnetic fields down to 50 mK is shown in **Figure 4a**. At zero field, the specific heat data exhibits two consecutive anomalies around $T_1$ = 0.11 K and $T_2$ = 2.81 K.

Specific heat capacity of fitted phonons ($C_{ph}$) was analyzed by a model of Debye and Einstein modes in the temperature regime from 30 to 100 K. The heat capacity formula of the Debye and Einstein model is expressed as a formula $C_{\mathrm{ph}} = 9N_{\mathrm{A}}k_{\mathrm{B}}c\left(\frac{T}{\Theta_{\mathrm{D}}}\right)^{3}\int_{0}^{\Theta_{\mathrm{D}}/T}\frac{x^{4}e^{x}}{(e^{x}-1)^{2}}dx + 3N_{\mathrm{A}}k_{\mathrm{B}}(1-c)\left(\frac{\Theta_{\mathrm{E}}}{T}\right)^{2}\frac{\exp(\frac{\Theta_{\mathrm{E}}}{T})}{\left[\exp(\frac{\Theta_{\mathrm{E}}}{T})-1\right]^{2}}$ , Where $N_A$ is Avogadro's number, $k_B$ is Boltzmann's constant, $\Theta_D$ and $\Theta_E$ are Debye and Einstein temperatures, and $c$ is a fitting coefficient.[36] By fitting the values of the fitting parameters, $c$ = 0.33, $\Theta_E$ = 550.11 K, and $\Theta_D$ = 224.60 K can be obtained (**Figure 4b**), respectively. The phonon-specific heat was obtained by extending the fitted curve from 0.05 to 290 K. However, above 140 K, the Debye and Einstein fits deviate from the experimental data. The magnetic specific heat capacity, $C_M$, is obtained by subtracting the $C_{ph}$ data from the experimental data at the corresponding temperatures. To ensure accurate calculation, the interpolation and extrapolation methods were employed.[37] The experimental value of magnetic entropy ($S_M$) was obtained by calculating using the formula $S_{\mathrm{M}} = \int_{0}^{T}(C_{\mathrm{M}}/T)dT$. The calculated results are shown in **Figure 4c**. It was observed that at 0 T, the magnetic specific heat almost dropped to zero at 30 K, and then resumed increasing with the rise in temperature at 140 K. This also led to that at low temperatures, the magnetic entropy exhibited a plateau close to $R\ln2$, indicating that the low-energy physics is governed by an effective quasi-doublet ($S_{eff}$ = 1/2) system derived from the crystal-field-split $J$ = 6 multiplet of $Tm^{3+}$. As temperatures increase further, the magnetic entropy progressively exceeds $R\ln 2$ and eventually approaches the full-multiplet limit of $R\ln(2J+1) = R\ln 13$ at high temperatures. This phenomenon is also observed in TMGO and originates from the pronounced CEF effects on the $Tm^{3+}$ ions.[10]

The magnetic specific heat in different fields is shown in **Figure 4d**. As the magnetic field increases, the anomaly at $T_1$ is gradually suppressed and nearly disappears around 1.5 T. Meanwhile, the anomaly at $T_2$ evolves into a sharp field-induced phase transition peak with increasing field. However, as the field further increases, this peak is subsequently suppressed and broadens once again at higher fields. At low temperatures, the system is dominated by quantum fluctuations arising from the pseudospin-1/2 quasi-doublet ground state on a geometrically frustrated TL. No conventional LRO is observed down to 50 mK under zero field, and the system exhibits two broad specific heat anomalies at $T_1$ and $T_2$. At zero field, the $T_2$ anomaly manifests as a broad hump, which may hint at the possible relevance of a BKT phase, similar to the parent compound TMGO.[27] This possibility is further corroborated by the low-temperature differential susceptibility $dM/dH$ for $B_\perp$, which follows a power-law behavior $dM/dH \sim H^{-\alpha}$ within the field range of 0.6–0.9 T with a fitted exponent $\alpha = 2/3$. Such nontrivial scaling behavior is suggestive of critical fluctuations associated with an emergent clock-like anisotropy and possible BKT physics in the TZGO (**Figure S3**).[27,38,39] With increasing magnetic field, the non-monotonic trajectory of the $T_2$ transition temperature outlines a characteristic dome-shaped phase boundary and provides evidence for the stabilization of the UUD spin configuration. As the field exceeds a critical value, the UUD order is completely suppressed, and the system is continuously driven into a field-polarized state.

## 3. Conclusion

In summary, we have synthesized millimeter-sized single crystals of TZGO using the high-temperature self-flux method. The dominant AFM interactions were evidenced by the large negative Weiss temperature, and significant easy-*c*-axis magnetic anisotropy was supported by the magnetization measurements. Under an external magnetic field, the system exhibits a clear one-third magnetization plateau, indicating a field-induced UUD (↑↑↓) spin configuration. The ultra-low-temperature specific heat measurement reveals anomalies at 0.11 K and 2.81 K, rather than the formation of conventional long-range magnetic order, highlighting the dominant role of strong spin fluctuations and suggesting the potential realization of a BKT phase. The above results reveal the interplay of geometric frustration, strong anisotropy, and non-Kramers doublet physics, and establish a Tm-based TL antiferromagnet as a platform for studying anisotropic frustrated magnetism and field-induced quantum magnetic phases.

## 4. Method

### 4.1. Single crystal growth

The single-crystal TZGO was synthesized by the high-temperature self-flux method. $Tm_2O_3$, $Ga_2O_3$, and ZnO were mixed in a ratio of 1:1:2 and thoroughly ground in an agate mortar. Subsequently, the mixture was fired in a high-temperature furnace using a crucible with a crucible lid. The sample was heated to 1570 ℃ and kept at this temperature for 1 day. Then, it was cooled to 1250 ℃ over 2 days and finally rapidly cooled to room temperature. The product was ultrasonically treated in deionized water for 1 hour, yielding transparent TZGO crystals (**Figure 1e**).

### 4.2. P-XRD

High-quality single crystals of TZGO were carefully selected under an optical microscope and thoroughly ground to obtain polycrystalline powder. Powder X-ray diffraction (P-XRD) measurement and the transparent millimeter-sized plate-like single crystal diffraction (SC-XRD) measurement were performed using a Rigaku SmartLab 9 kW diffractometer. The experiment was conducted in the Bragg-Brentano geometry using Cu $K_{\alpha}$ radiation ($K_{\alpha_1}$ = 1.54056 Å, $K_{\alpha_2}$ = 1.54439 Å) as the X-ray source. Data were collected over a $2\theta$ range of 5° to 110°, with a step size of 0.02° and a scanning rate of 2° per minute for polycrystalline powder and over a $2\theta$ range of 10° to 80°, with a step size of 0.02° and a scanning rate of 10° per minute for plate-like single crystal. The obtained polycrystalline powder diffraction data were refined using the GSAS-II software package[33].

**4.3. SC-XRD**

SC-XRD measurements were carried out at 300 K using a Bruker D8 VENTURE diffractometer equipped with a PHOTON III CPAD detector. Monochromatic Mo $K_{\alpha}$ radiation ($\lambda$ = 0.71073 Å), produced via a graphite monochromator, served as the X-ray source. Raw data were processed using APEX4 software, including integration, scale, absorption correction, and space group determination. Additional refinement of the crystal structure was performed using the ShelXL least-squares algorithm implemented in Olex2.[40,41] Synthetic precession images were generated from raw data using APEX4 and simulated using SingleCrystal software based on the CIF data for TZGO.[34]

**4.4. Magnetic measurement**

Magnetic susceptibility ($\chi$) measurements were performed using Physical Property Measurement System (PPMS) and Magnetic Property Measurement System (MPMS3) with $^3$He option manufactured by Quantum Design. Magnetic measurements were performed with the field applied (i) perpendicular to the *a* axis, i.e., within the TL plane ($B_{\parallel}$), and (ii) along the *c* axis, i.e., perpendicular to the TL plane ($B_{\perp}$). Zero-field-cooling (ZFC) susceptibility measurements were extended up from 1.8 to 100 K under 0.2 T field for $B_{\perp}$ and $B_{\parallel}$, respectively. The field dependence of magnetization was also measured at various temperatures from 0 to 7 T. In addition, the ZFC curve from 0.4 to 2 K and the field dependence of the magnetization intensity from 0 to 7 T were measured using the MPMS3 with a $^3$He option for $B_{\perp}$ and $B_{\parallel}$.

**4.5. Specific heat measurement**

Specific heat measurements were performed using the PPMS at 1.8−300 K under an applied magnetic field ranging from 0 to 9 T, with the field applied along the $B_{\perp}$ direction. The background heat capacity of the N-grease was measured separately before the sample measurement and subtracted from the sample data to correct for the grease contribution. Data from 50 mK to 2 K were measured using the PPMS with the Dilution Refrigerator (DR) option. Taking into account the quality error, the data from the DR option measurement was calibrated using the standard specific heat data above 1.8 K.

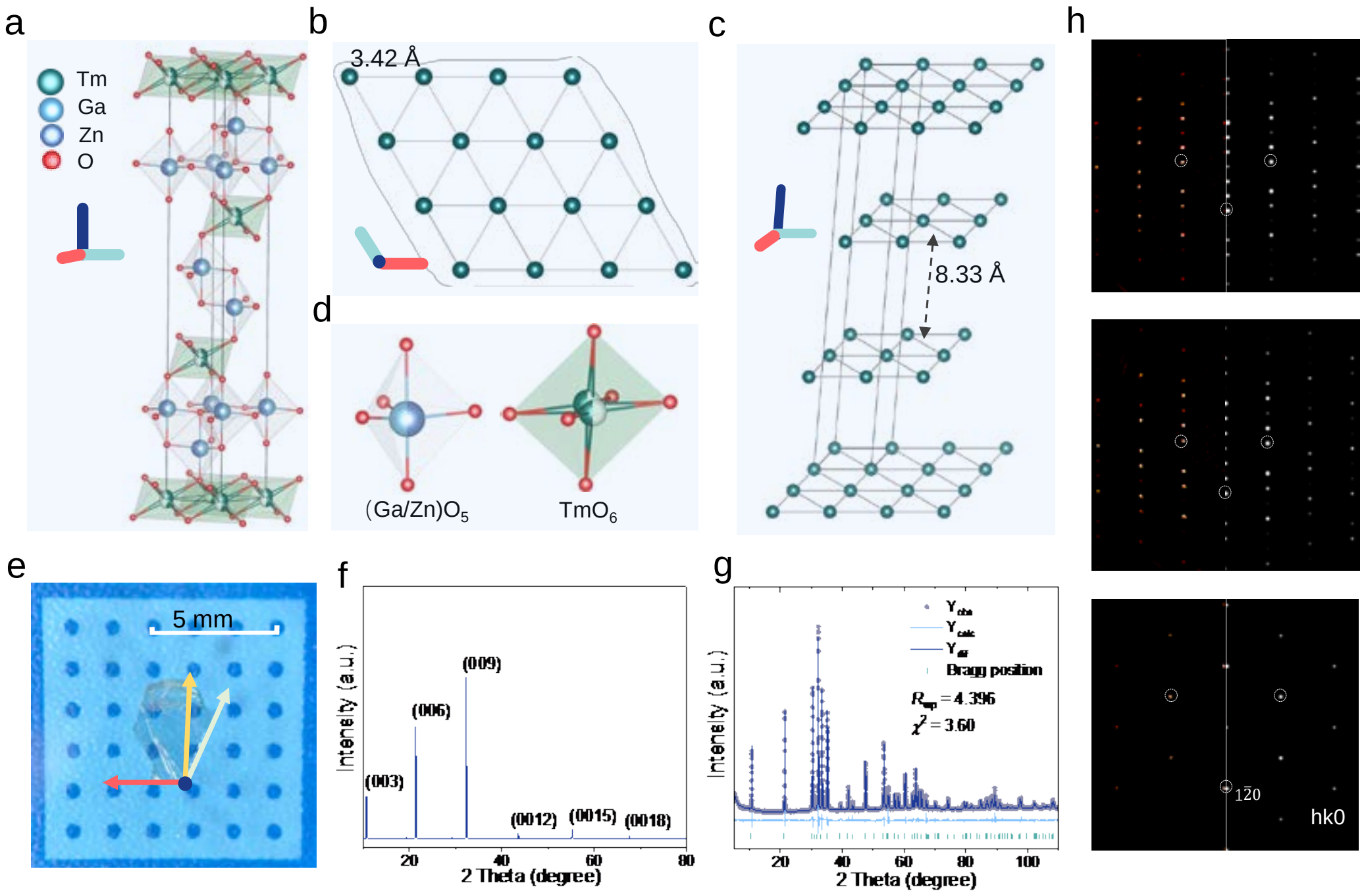

**Figure 1. Crystal structure, morphology, and structural characterization of TZGO.** (a) Unit cell. (b) Tm-based TL plane. (c) Interlayer magnetic framework. (d) Triangular bipyramid for $(Ga/Zn)O_5$ and octahedra for $TmO_6$. (e) The picture of a single crystal. (f) P-XRD pattern of crystals on the (00l) direction. (g) Rietveld refinement of P-XRD patterns. (h) Diffracted reflections in reciprocal lattice planes (0kl), (h0l), and (hk0) determined from experimental (left) and calculated (right) data, respectively (shown symmetrically).

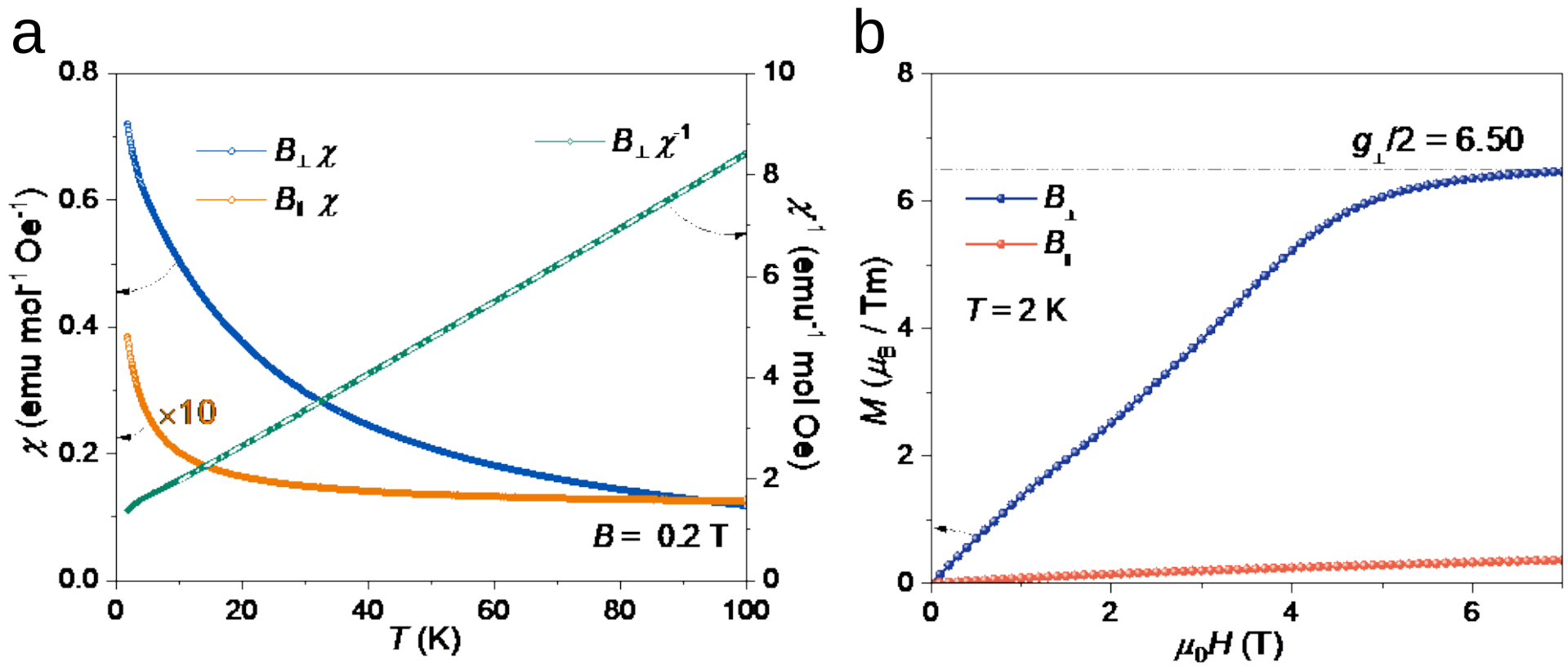

**Figure 2. Anisotropic Magnetism above 2 K.** (a) Temperature-dependent magnetic susceptibility $\chi$, and inverse susceptibility $\chi^{-1}$, with the external field $B$ = 0.2 T for $B_{\perp}$. And temperature-dependent magnetic susceptibility $\chi$ for $B_{\parallel}$ (To facilitate an intuitive presentation, it has been multiplied by a multiple of 10). (b) The variation curve of magnetization intensity $M$ with the external field for $B_{\parallel}$ and $B_{\perp}$ when $T$ = 2 K.

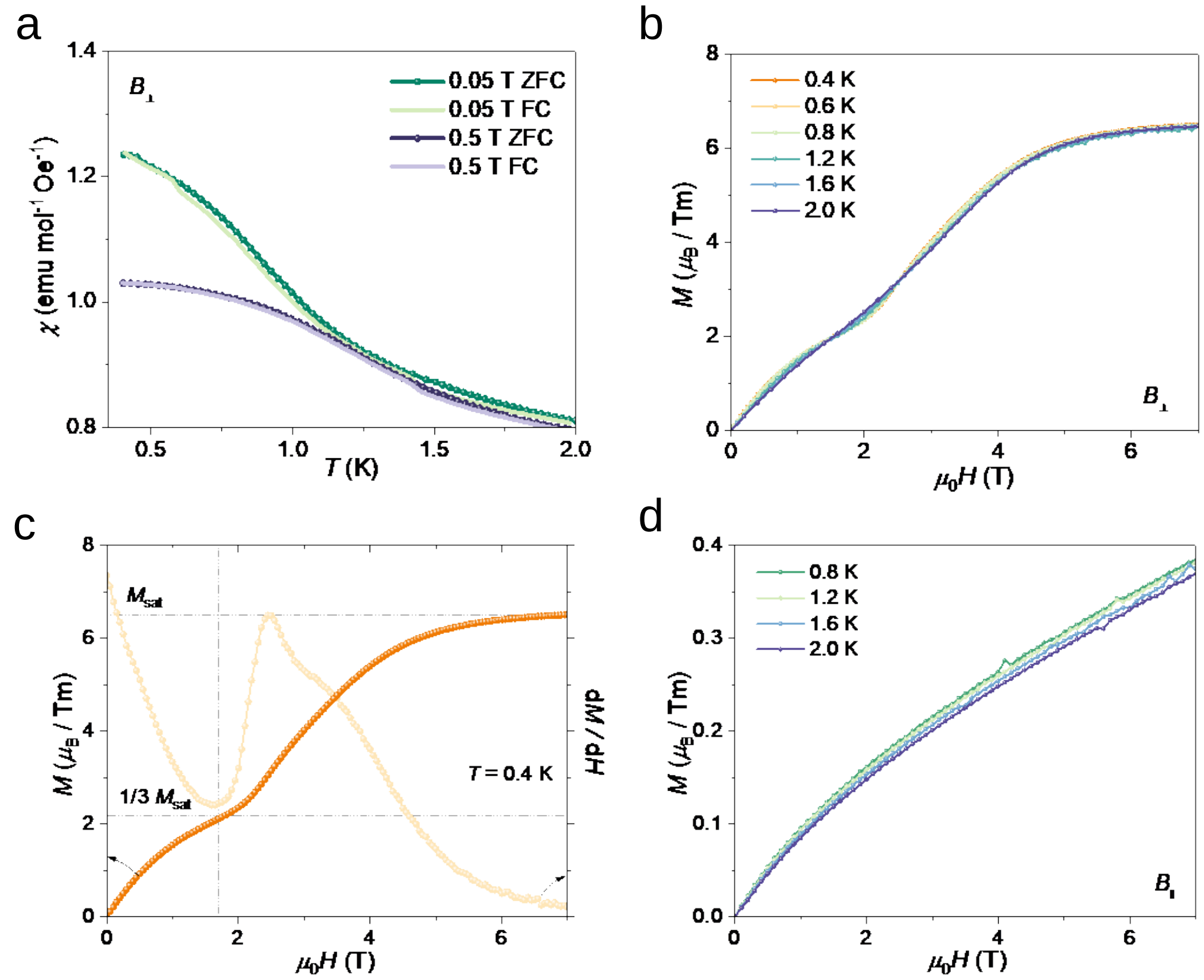

**Figure 3. Low-temperature Magnetism.** (a) Temperature-dependent magnetic susceptibility $\chi$ below 2 K for $B_\perp$. (b) Field dependence of magnetization $M(B)$ at various temperatures for $B_\perp$. (c) Magnetization $M(H)$ and its differential d$M$/d$H$ at $T$ = 0.4 K for $B_\perp$. (d) Field dependence of magnetization $M(H)$ at different temperatures for $B_\parallel$.

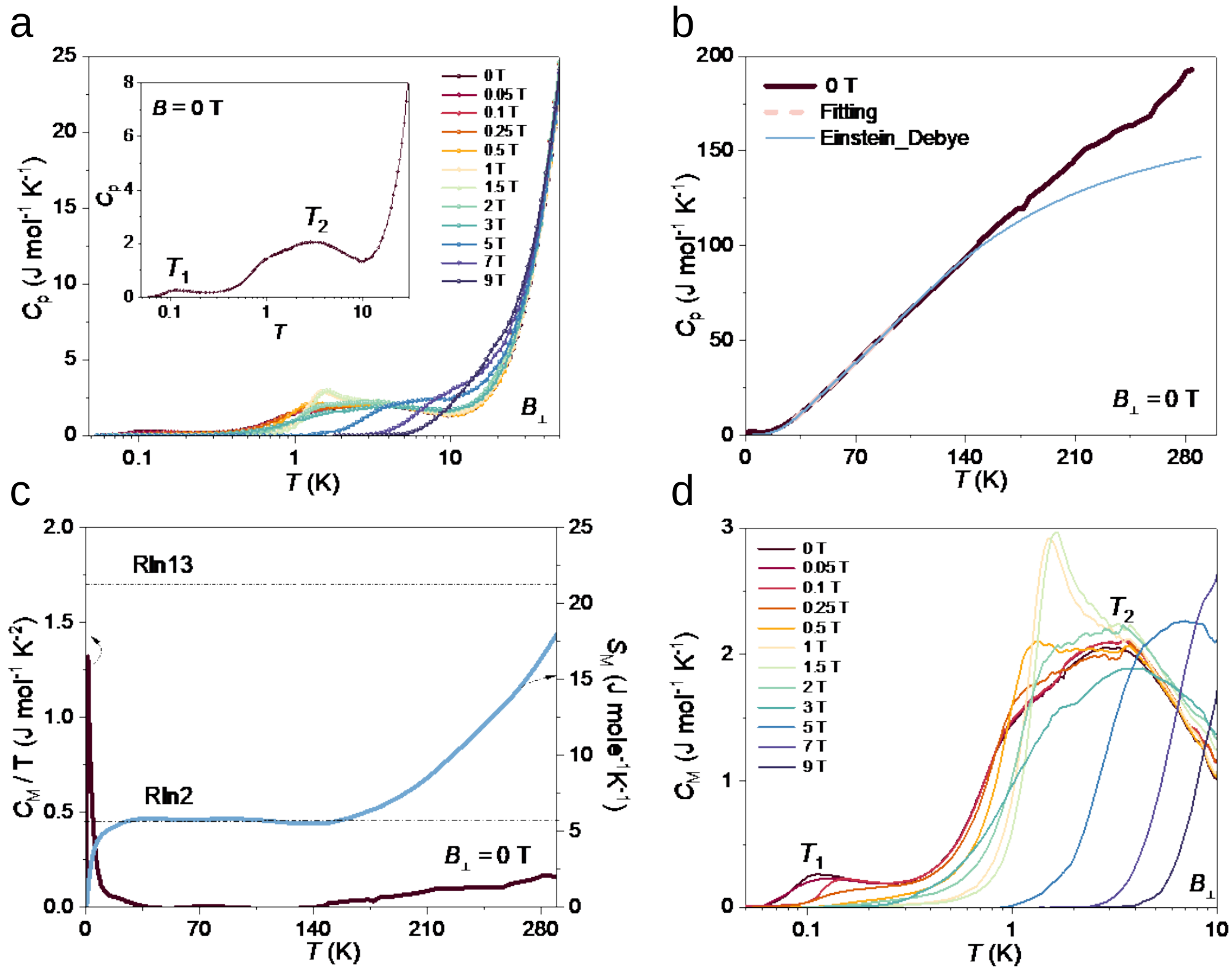

**Figure 4. Thermodynamic properties.** (a) The temperature dependence of specific heat data $C_p(T)$ under different magnetic fields. The inset highlights the zero-field data, with two broad maxima clearly visible for better visualization. (b) Zero-field specific heat $C_p(T)$ from 290 K to 0.05 K. The pink dashed line shows the phonon contribution fitted using the Einstein-Debye model in the range 30–100 K, and the blue solid line shows the extrapolated fit. (c) Magnetic specific heat, $C_M/T$, and the magnetic entropy ($S_M$) in zero field. (d) Temperature dependence of $C_M$ under different magnetic fields.

### Author contributions

Corresponding Author

*Shu Guo – Shenzhen Institute for Quantum Science and Engineering, Southern University of Science and Technology, Shenzhen 518055, China; International Quantum Academy, Shenzhen 518048, China; orcid.org/0000-0002-2098-8904; Email: guos@sustech.edu.cn/shu-guo@outlook.com.

Authors

Zhaoyi Li – School of Physics and Electronics, Henan University, Kaifeng 475004, China; Shenzhen Institute for Quantum Science and Engineering, Southern University of Science and Technology, Shenzhen 518055, China; International Quantum Academy, Shenzhen 518048, China; orcid.org/0009-0001-4774-6307.

Qinchen Duan – Tsung-Dao Lee Institute, Shanghai Jiao Tong University, Shanghai 201210, People’s Republic of China; School of Physics and Astronomy, Shanghai Jiao Tong University, Shanghai 200240, People’s Republic of China.

Bo Wen – School of Physics and Electronics, Henan University, Kaifeng 475004, China; orcid.org/0000-0001-7252-918X.

Ruidan Zhong – Tsung-Dao Lee Institute, Shanghai Jiao Tong University, Shanghai 201210, People’s Republic of China; School of Physics and Astronomy, Shanghai Jiao Tong University, Shanghai 200240, People’s Republic of China.

Author Contributions

S. G. conceived and directed this project. Z. L., Q. D., R. Z., and S. G. conducted the main experimental studies. Z. L. and S. G. analyzed and interpreted the experimental data. All authors contributed to the writing and editing of the manuscript and agreed on the final version.

### Notes

The authors declare no competing interests.

### Supporting Information

Crystal structures, relevant refinement details, crystallographic information, and additional magnetic characterizations.

The Cambridge Crystallographic Data Centre number 2552441 contains supplementary crystallographic data for TZGO. This data can be obtained free of charge via www.ccdc.cam.ac.uk/data_request/cif, by emailing data_request@ccdc.cam.ac.uk, or by contacting The Cambridge Crystallographic Data Centre, 12 Union Road, Cambridge CB2 1EZ, UK; fax: + 441223336033.

**Acknowledgments**

The work was supported by the National Key R&D of China under 2022YFA1402702 and 2021YFA1401600, the Guangdong Pearl River Talent Plan (2023QN10C793), and the National Natural Science Foundation of China with Grants Nos. 22205091, 12334008, and 12374148.

**Data Availability Statement**

The data supporting the findings of this study are available from the corresponding author upon reasonable request.

# Supporting Information

**Table of Contents**

**Figure S1.** Temperature-dependent magnetic susceptibility $\chi$ and inverse susceptibility $\chi^{-1}$with the external field $B$ = 0.2 T for $B_{\parallel}$.

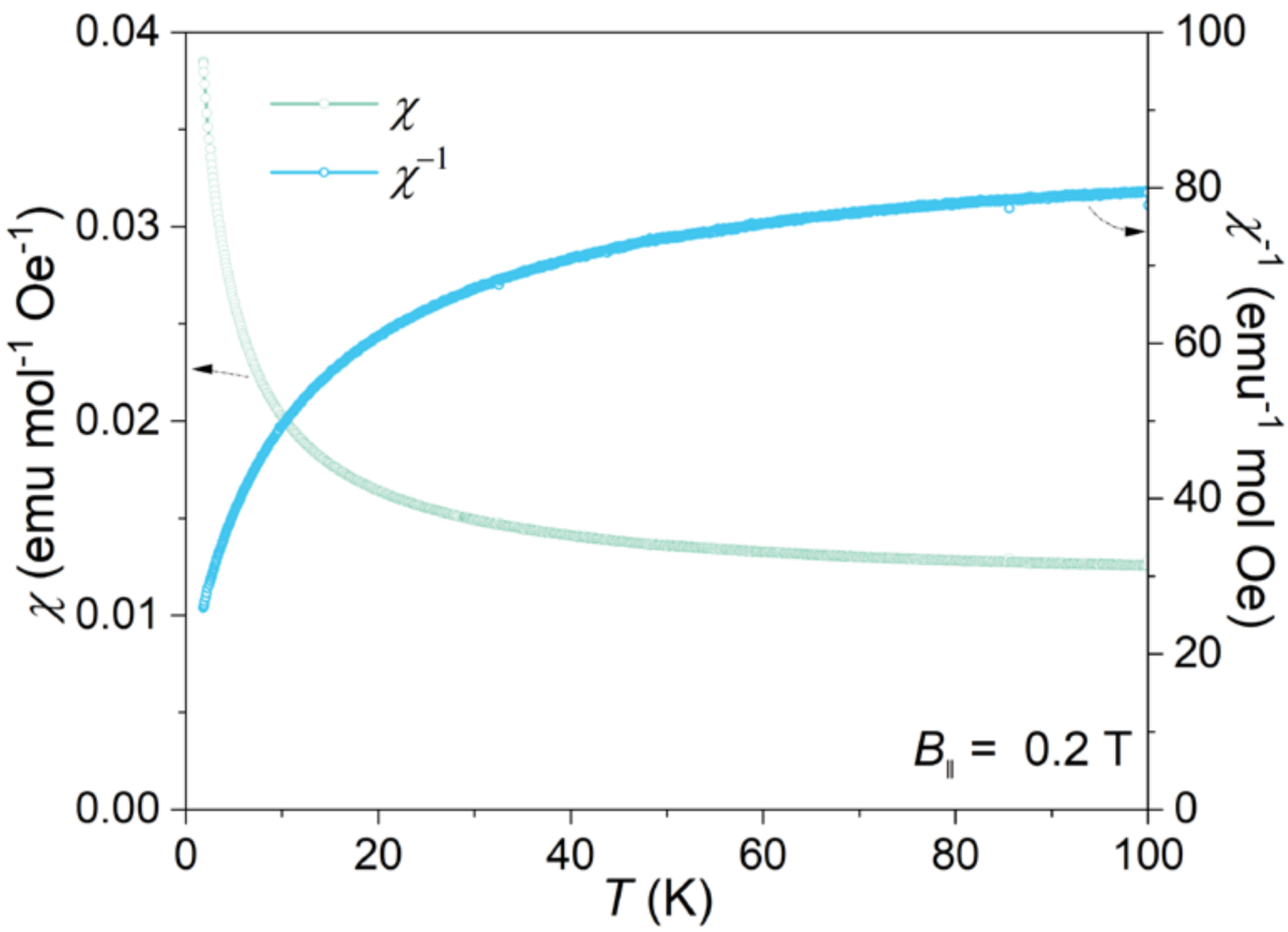

**Figure S2.** Temperature-dependent magnetic susceptibility $\chi$ with the external field for $B_{\parallel}$ below 2 K.

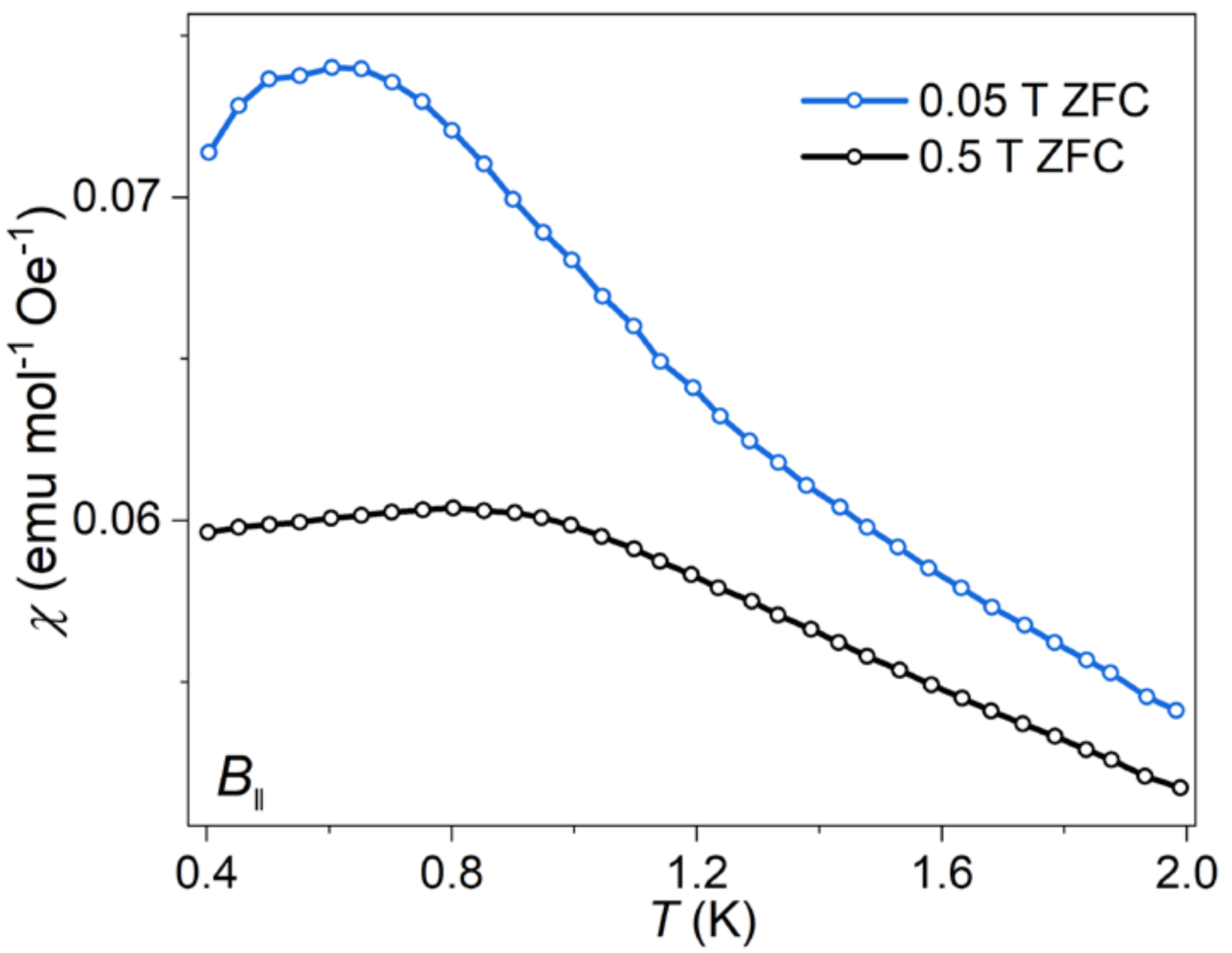

**Figure S3.** (a) The differential susceptibility d$M$/d$H$ for $B_{\perp}$ at different temperatures. (b) Fits of the differential susceptibility to the power-law function d$M$/d$H$ ~ $H^{-\alpha}$ with α = 2/3 below 1 K.

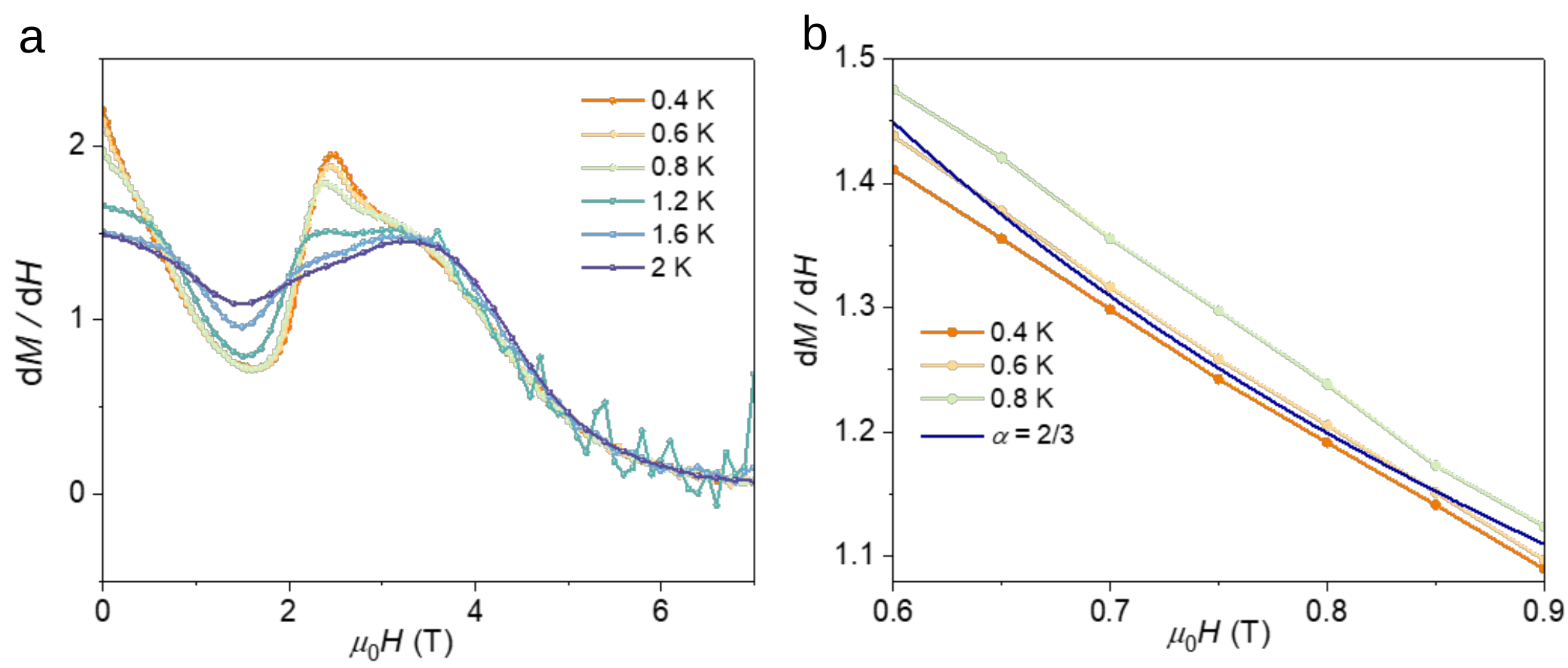

**Table S1.** Crystal Data and Structure Refinements for TZGO.

| | |
|---|---|
| Empirical formula | $TmZnGaO_4$ |
| Temperature/K | 301(2) K |
| Crystal system | Trigonal |
| Space group | *R*-3*m* |
| Formula weight | 368.03 |
| *a*/Å | 3.4251 (7) |
| *c*/Å | 24.992 (11) |
| Volume/$Å^3$ | 253.91 (16) |
| *Z* | 3 |
| $\rho_{calc}$/g/$cm^3$ | 7.220 |
| $\mu$/$mm^{-1}$ | 40.750 |
| *F*(000) | 486 |
| Crystal size/$mm^3$ | 0.029 × 0.025 × 0.025 |
| Radiation | Mo $K_{\alpha}$ ($\lambda$ = 0.71073) |
| $\theta$ range for data collection/° | 2.445-27.349 |
| Index ranges | $-4\leq h \leq4, -4\leq k \leq4, -31\leq l \leq31$ |
| Goodness-of-fit on $F^2$ | 1.180 |
| max density difference/e/$Å^3$ | 1.943 |
| min density difference/e/$Å^3$ | -1.111 |
| Final *R* indexes [$I \geq 2\sigma(I)$] | $R_1$ = 0.0254, $wR_2$ = 0.0572 |
| Final *R* indexes [all data] | $R_1$ = 0.0266, $wR_2$ = 0.0578 |

**Table S2.** Fractional atomic coordinates and equivalent isotropic displacement parameters (Å$^2$) for TZGO.

| TZGO | | | | | |
|---|---|---|---|---|---|
| | *x* | *y* | *z* | $U_{eq}$ | Occ. |
| Tm1 | 0.666667 | 0.333333 | 0.3270 (3) | 0.0086 (13) | 0.5 |
| Ga1 | 0.000000 | 0.000000 | 0.21458 (5) | 0.0095 (5) | 0.5 |
| Zn1 | 0.000000 | 0.000000 | 0.21458 (5) | 0.0095 (5) | 0.5 |
| O1 | 0.666667 | 0.333333 | 0.2045 (5) | 0.023 (2) | |
| O2 | 0.000000 | 0.000000 | 0.2906 (3) | 0.0088 (18) | |

**Table S3.** Atomic displacement parameters (Å$^2$) for TZGO.

| TZGO | | | | | |
|---|---|---|---|---|---|
| | $U$11 | $U$22 | $U$33 | $U$12 | $U$13 |
| Tm1 | 0.0045 (5) | 0.0045 (5) | 0.017 (4) | 0.0023 (2) | 0.000 |
| Ga1 | 0.0080 (6) | 0.0080 (6) | 0.0127 (9) | 0.0040 (3) | 0.000 |
| Zn1 | 0.0080 (6) | 0.0080 (6) | 0.0127 (9) | 0.0040 (3) | 0.000 |
| O1 | 0.020 (4) | 0.020 (4) | 0.028 (5) | 0.0101 (18) | 0.000 |
| O2 | 0.009 (3) | 0.009 (3) | 0.008 (4) | 0.0047 (13) | 0.000 |

**Table S4.** Geometric parameters (Å, °) for TZGO.

| TZGO | | | |
|---|---|---|---|
| Tm1—Tm1$^{i}$ | 0.316 (16) | Tm1—O2 | 2.177 (5) |
| Tm1—Tm1$^{ii}$ | 3.4251 (7) | Tm1—O2$^{ix}$ | 2.177 (5) |
| Tm1—Tm1$^{iii}$ | 3.4251 (7) | Tm1—O2$^{x}$ | 2.177 (5) |
| Tm1—Tm1$^{iv}$ | 3.4251 (7) | Ga1—O1$^{xi}$ | 2.144 (12) |
| Tm1—Tm1$^{v}$ | 3.4251 (7) | Ga1—O1 | 1.9933 (16) |
| Tm1—Tm1$^{vi}$ | 3.4251 (7) | Ga1—O1$^{ii}$ | 1.9934 (16) |
| Tm1—O2$^{vii}$ | 2.326 (6) | Ga1—O1$^{iii}$ | 1.9933 (16) |
| Tm1—O2$^{i}$ | 2.326 (6) | Ga1—O2 | 1.900 (9) |
| Tm1—O2$^{viii}$ | 2.326 (6) | | |
| Tm1$^{i}$—Tm1—Tm1$^{vi}$ | 90.000 (4) | O2$^{ix}$—Tm1—O2$^{vii}$ | 172.9 (4) |
| Tm1$^{iv}$—Tm1—Tm1$^{v}$ | 60 | O2—Tm1—O2$^{viii}$ | 80.4 (2) |
| Tm1$^{v}$—Tm1—Tm1$^{iii}$ | 60 | O2$^{viii}$—Tm1—O2$^{vii}$ | 94.8 (3) |
| Tm1$^{v}$—Tm1—Tm1$^{ii}$ | 120 | O2$^{ix}$—Tm1—O2$^{viii}$ | 80.4 (2) |
| Tm1$^{vi}$—Tm1—Tm1$^{iii}$ | 180 | O2$^{i}$—Tm1—O2$^{vii}$ | 94.8 (3) |
| Tm1$^{i}$—Tm1—Tm1$^{iv}$ | 90.001 (7) | Tm1$^{ii}$—Ga1—Tm1 | 59.79 (13) |
| Tm1$^{vi}$—Tm1—Tm1$^{v}$ | 120.000 (1) | Tm1$^{iii}$—Ga1—Tm1$^{ii}$ | 59.79 (13) |
| Tm1$^{vi}$—Tm1—Tm1$^{iv}$ | 60.000 (1) | Tm1$^{iii}$—Ga1—Tm1 | 59.79 (13) |
| Tm1$^{i}$—Tm1—Tm1$^{iii}$ | 90 | O1$^{iii}$—Ga1—Tm1$^{ii}$ | 112.9 (3) |
| Tm1$^{i}$—Tm1—Tm1$^{v}$ | 90.001 (3) | O1$^{xi}$—Ga1—Tm1 | 144.87 (8) |
| Tm1$^{iv}$—Tm1—Tm1$^{iii}$ | 120.000 (1) | O1—Ga1—Tm1 | 62.1 (3) |
| Tm1$^{vi}$—Tm1—Tm1$^{ii}$ | 120 | O1—Ga1—Tm1$^{ii}$ | 112.9 (3) |
| Tm1$^{i}$—Tm1—Tm1$^{ii}$ | 90 | O1$^{iii}$—Ga1—Tm1 | 112.9 (3) |
| Tm1$^{iii}$—Tm1—Tm1$^{ii}$ | 60 | O1$^{ii}$—Ga1—Tm1$^{ii}$ | 62.1 (3) |
| Tm1$^{iv}$—Tm1—Tm1$^{ii}$ | 180 | O1$^{ii}$—Ga1—Tm1 | 112.9 (3) |
| Tm1$^{i}$—Tm1—O2$^{i}$ | 58.2 (2) | O1—Ga1—Tm1$^{iii}$ | 112.9 (3) |
| Tm1$^{i}$—Tm1—O2$^{ix}$ | 114.7 (3) | O1$^{xi}$—Ga1—Tm1$^{iii}$ | 144.86 (8) |
| Tm1$^{i}$—Tm1—O2 | 114.7 (3) | O1$^{ii}$—Ga1—Tm1$^{iii}$ | 112.9 (3) |
| Tm1$^{i}$—Tm1—O2$^{viii}$ | 58.2 (2) | O1$^{iii}$—Ga1—Tm1$^{iii}$ | 62.1 (3) |
| Tm1$^{i}$—Tm1—O2$^{vii}$ | 58.2 (2) | O1$^{xi}$—Ga1—Tm1$^{ii}$ | 144.87 (8) |
| Tm1$^{i}$—Tm1—O2$^{x}$ | 114.7 (3) | O1$^{iii}$—Ga1—O1$^{ii}$ | 118.44 (15) |
| O2$^{x}$—Tm1—Tm1$^{iv}$ | 38.11 (17) | O1$^{iii}$—Ga1—O1 | 118.44 (15) |
| O2$^{i}$—Tm1—Tm1$^{vi}$ | 42.59 (15) | O1—Ga1—O1$^{xi}$ | 82.8 (3) |
| O2$^{x}$—Tm1—Tm1$^{iii}$ | 90 | O1—Ga1—O1$^{ii}$ | 118.44 (15) |
| O2—Tm1—Tm1$^{v}$ | 90 | O1$^{iii}$—Ga1—O1$^{xi}$ | 82.8 (3) |
| O2—Tm1—Tm1$^{iii}$ | 38.11 (17) | O1$^{ii}$—Ga1—O1$^{xi}$ | 82.8 (3) |
| O2$^{i}$—Tm1—Tm1$^{iv}$ | 42.59 (15) | O2—Ga1—Tm1 | 35.13 (8) |
| O2$^{viii}$—Tm1—Tm1$^{iii}$ | 90 | O2—Ga1—Tm1$^{iii}$ | 35.13 (8) |
| O2$^{viii}$—Tm1—Tm1$^{iv}$ | 137.41 (15) | O2—Ga1—Tm1$^{ii}$ | 35.13 (8) |
| O2$^{ix}$—Tm1—Tm1$^{iv}$ | 90.000 (1) | O2—Ga1—O1$^{ii}$ | 97.2 (3) |
| O2$^{i}$—Tm1—Tm1$^{v}$ | 90 | O2—Ga1—O1$^{iii}$ | 97.2 (3) |
| O2$^{ix}$—Tm1—Tm1$^{iii}$ | 141.88 (17) | O2—Ga1—O1$^{xi}$ | 180 |
| O2$^{viii}$—Tm1—Tm1$^{v}$ | 137.41 (15) | O2—Ga1—O1 | 97.2 (3) |
| O2$^{i}$—Tm1—Tm1$^{iii}$ | 137.41 (15) | Ga1$^{ix}$—O1—Ga1 | 118.44 (15) |
| O2$^{vii}$—Tm1—Tm1$^{iv}$ | 90.000 (1) | Ga1$^{ix}$—O1—Ga1$^{x}$ | 118.44 (15) |
| O2$^{vii}$—Tm1—Tm1$^{iii}$ | 42.59 (15) | Ga1—O1—Ga1$^{xi}$ | 97.2 (3) |
| O2—Tm1—Tm1$^{ii}$ | 38.11 (18) | Ga1$^{x}$—O1—Ga1 | 118.44 (15) |
| O2$^{viii}$—Tm1—Tm1$^{vi}$ | 90 | Ga1$^{x}$—O1—Ga1$^{xi}$ | 97.2 (3) |
| O2$^{x}$—Tm1—Tm1$^{ii}$ | 141.88 (17) | Ga1$^{ix}$—O1—Ga1$^{xi}$ | 97.2 (3) |
| O2$^{vii}$—Tm1—Tm1$^{ii}$ | 90.000 (1) | Tm1$^{i}$—O2—Tm1$^{vii}$ | 94.8 (3) |
| O2$^{vii}$—Tm1—Tm1$^{v}$ | 42.59 (15) | Tm1$^{ii}$—O2—Tm1$^{viii}$ | 7.1 (4) |
| O2$^{ix}$—Tm1—Tm1$^{ii}$ | 90.000 (1) | Tm1$^{ii}$—O2—Tm1$^{vii}$ | 99.6 (2) |
| O2—Tm1—Tm1$^{vi}$ | 141.89 (17) | Tm1$^{iii}$—O2—Tm1$^{ii}$ | 103.8 (3) |
| O2$^{i}$—Tm1—Tm1$^{ii}$ | 137.41 (15) | Tm1$^{i}$—O2—Tm1$^{viii}$ | 94.8 (3) |
| O2$^{ix}$—Tm1—Tm1$^{vi}$ | 38.11 (17) | Tm1$^{iii}$—O2—Tm1 | 103.8 (3) |
| O2—Tm1—Tm1$^{iv}$ | 141.89 (17) | Tm1$^{iii}$—O2—Tm1$^{viii}$ | 99.6 (2) |
| O2$^{x}$—Tm1—Tm1$^{vi}$ | 90 | Tm1$^{ii}$—O2—Tm1 | 103.8 (3) |
| O2$^{ix}$—Tm1—Tm1$^{v}$ | 141.88 (17) | Tm1—O2—Tm1$^{viii}$ | 99.6 (2) |
| O2$^{vii}$—Tm1—Tm1$^{vi}$ | 137.41 (15) | Tm1$^{iii}$—O2—Tm1$^{i}$ | 99.6 (2) |
| O2$^{x}$—Tm1—Tm1$^{v}$ | 38.11 (17) | Tm1$^{iii}$—O2—Tm1$^{vii}$ | 7.1 (4) |
| O2$^{viii}$—Tm1—Tm1$^{ii}$ | 42.59 (15) | Tm1$^{ii}$—O2—Tm1$^{i}$ | 99.6 (2) |
| O2—Tm1—O2$^{ix}$ | 103.8 (3) | Tm1—O2—Tm1$^{vii}$ | 99.6 (2) |
| O2$^{i}$—Tm1—O2$^{viii}$ | 94.8 (3) | Tm1—O2—Tm1$^{i}$ | 7.1 (4) |
| O2$^{x}$—Tm1—O2$^{vii}$ | 80.4 (2) | Tm1$^{viii}$—O2—Tm1$^{vii}$ | 94.8 (3) |
| O2—Tm1—O2$^{i}$ | 172.9 (4) | Ga1—O2—Tm1$^{vii}$ | 121.8 (2) |
| O2—Tm1—O2$^{vii}$ | 80.4 (2) | Ga1—O2—Tm1$^{viii}$ | 121.8 (2) |
| O2$^{ix}$—Tm1—O2$^{i}$ | 80.4 (2) | Ga1—O2—Tm1$^{i}$ | 121.8 (2) |
| O2$^{x}$—Tm1—O2$^{viii}$ | 172.9 (4) | Ga1—O2—Tm1$^{ii}$ | 114.7 (3) |
| O2$^{x}$—Tm1—O2$^{i}$ | 80.4 (2) | Ga1—O2—Tm1 | 114.7 (3) |
| O2$^{ix}$—Tm1—O2$^{x}$ | 103.8 (3) | Ga1—O2—Tm1$^{iii}$ | 114.7 (3) |
| O2—Tm1—O2$^{x}$ | 103.8 (3) | | |

Symmetry codes: (i) $-x+4/3, -y+2/3, -z+2/3$; (ii) $x-1, y-1, z$; (iii) $x-1, y, z$; (iv) $-x+y+2, -x+2, z$; (v) $x, y+1, z$; (vi) $-x+y+2, -x+1, z$; (vii) $-x+1/3, -y+2/3, -z+2/3$; (viii) $-x+1/3, -y-1/3, -z+2/3$; (ix) $x+1, y, z$; (x) $x+1, y+1, z$; (xi) $-x+2/3, -y+1/3, -z+1/3$.